\documentclass{article}
\usepackage[dvips]{graphicx}  

\begin{document}
\newcommand{\rum}{\rule{0.5pt}{0pt}}
\newcommand{\rub}{\rule{1pt}{0pt}}
\newcommand{\rim}{\rule{0.3pt}{0pt}}
\newcommand{\numtimes}{\mbox{\raisebox{1.5pt}{${\scriptscriptstyle \times}$}}}
\renewcommand{\refname}{References}


\begin{center}
{\Large\bf  The Roland De Witte 1991 Detection of Absolute Motion and Gravitational Waves
\rule{0pt}{13pt}}\par

\bigskip

Reginald T. Cahill \\ 

{\small\it School of Chemistry, Physics and Earth Sciences, Flinders University,

Adelaide 5001, Australia\rule{0pt}{13pt}}\\

\raisebox{-1pt}{\footnotesize E-mail: Reg.Cahill@flinders.edu.aul}\par

\bigskip\smallskip

Published: Progress in Physics {\bf 3},  60-65, 2006. \\ 

\vspace{3mm}

{\small\parbox{11cm}{%
In 1991 Roland De Witte carried out an experiment in Brussels in which variations in  the one-way speed of RF waves through a coaxial cable were recorded over 178 days. The data from  this experiment shows that De Witte had detected absolute motion of the earth through space, as had six earlier experiments, beginning with the Michelson-Morley experiment of 1887. His results are in excellent agreement with the extensive data from the Miller  1925/26 detection of absolute motion using a gas-mode Michelson interferometer atop Mt.\hspace{-0.4mm}Wilson, California.  The De Witte data  reveals turbulence in the flow which amounted to the detection of gravitational waves. Similar effects were also seen by Miller, and by Torr and Kolen in their coaxial cable experiment. Here we bring together what is known about the De Witte experiment.

\rule[0pt]{0pt}{0pt}}}\bigskip

\end{center}

\section{Introduction\label{section:introduction}}

\begin{figure}[h]
\hspace{60mm}\includegraphics[scale=0.75]{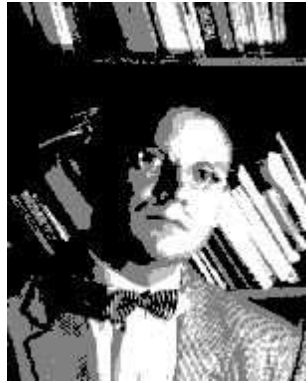}
\caption{\small{ Roland De Witte. }
\label{fig:DeWittePhoto}}\end{figure}

Ever since the 1887 Michelson-Morley experiment \cite{MM}  to detect absolute motion, that is motion relative to space, by means of the anisotropy of the speed of light,  physicists in the main have believed that such absolute motion was unobservable, and even meaningless\footnote{The older terminology was that of detecting motion relative to an  {\it ether} that was embedded in a geometrical space.  However the more modern understanding does away with both the ether and a geometrical space, and uses a structured dynamical {\it 3-space}, as in \cite{Book, AMGE}. }. This was so  after Einstein proposed as one of his postulates for his Special Theory of Relativity that the speed of light was the same for all observers, that it was necessarily isotropic. This was despite the fact that the Michelson-Morley experiment did observe fringe shifts of the form indicative of such an anisotropy.  The whole issue has been one of great confusion over the last 100 years or so.  This confusion arose from deep misunderstandings of the theoretical structure of Special Relativity, but also because  ongoing detections of the anisotropy of the speed of light were treated with contempt, rather than being rationally discussed.    The intrinsic problem all along has been that the observed anisotropy of the speed of light also affects the very apparatus being used to measure the anisotropy. In particular the Lorentz-Fitzgerald length contraction effect must be included in the analysis of the interferometer when the calibration constant for the device is calculated. The calibration constant determines what value of the speed of light anisotropy is to be determined from an observed fringe shift as the apparatus is rotated. Only in 2002 was it discovered that the calibration constant is very much smaller  than had been assumed \cite{MMCK, MMC}, and that the observed fringe shifts corresponded to a speed in excess of 0.1\% of the speed of light.  That discovery showed that the presence of a gas in the light path is essential if the interferometer is to act as a detector of absolute motion, and that a vacuum operated interferometer is totally incapable of detecting absolute motion. That physics has suppressed this effect for over 100 years is a major indictment of physics.   There have been in all seven detections of such anisotropy, with five being Michelson interferometer experiments \cite{MM, Miller, C5, C6, C7}, and two being one-way RF coaxial cable propagation time experiments, see \cite{Book, AMGE} for extensive discussion and analysis of the experimental data.  The most thorough interferometer experiment was by Miller in 1925/26.  He accumulated sufficient data  that in conjunction with  the new calibration understanding, the velocity of motion of the solar system could be determined\footnote{There is a possibility that the direction is opposite to this direction} as ($\alpha =5.2^{hr}$, $\delta =-67^0$), with a speed of  $420\pm 30$km/s. This local (in the galactic sense) absolute motion is different from the Cosmic Microwave Background (CMB) anisotropy  determined motion, in the direction ($\alpha=11.20^{hr}, \delta=-7.22^0$) with speed  $369$km/s; this is motion relative to the source of the CMB, namely  relative to the distant universe.

The first one-way  coaxial cable  speed-of-propagation experiment was
performed at the Utah University in 1981 by Torr and  Kolen \cite{Torr}. This involved two rubidium vapor clocks placed approximately 500m apart with a 5 MHz sinewave RF signal propagating between the clocks via a buried nitrogen filled coaxial cable maintained at a constant pressure of $\sim$2 psi.  There is no reference to Miller's result in the Torr and  Kolen paper.  There is a   projection of the absolute motion velocity onto the East-West cable and Torr and Kolen did observe an effect in that, while the round speed time remained constant within 0.0001\%c, variations in the one-way travel time  were observed. The maximum effect occurred, typically,  at the  times predicted using the Miller velocity \cite{Book, AMGE}.   So the results of this experiment are also in remarkable agreement with the Miller direction, and the speed of 420 km/s. As well Torr and Kolen reported fluctuations in both the magnitude, from 1 - 3 ns, and the time of  maximum variations in travel time.

However during 1991 Roland De Witte performed the  most extensive RF travel time experiment, accumulating data over 178 days . His data is in complete agreement with the 1925/26 Miller experiment.   These two experiments will eventually be recognised as two of the most significant experiments in physics, for independently and using different experimental techniques they detected the same velocity of absolute motion.  But also they detected turbulence in the flow of space past the earth; non other than gravitational waves.  Both Miller and De Witte have been repeatably  attacked for their discoveries.  Of course all seven experiments indicate that the Einstein postulate regarding the anisotropy of the speed of light is falsified, but that is not in conflict with the confirmed correctness of various so-called relativistic effects, rather it indicates that these effects are to be understood as being caused by absolute motion of systems relative to space, as suggested by Lorentz in the 19th century. So it turns out that the evidence from more than 100 years has been that Lorentz relativity is correct, and that the Einstein relativity is falsified.  While Miller was able to publish his results \cite{Miller}, and indeed the original data sheets were recently discovered at Case Western Reserve University, Cleveland, Ohio, De Witte was never  permitted to publish his data in a physics journal. The only source of his data was from a e-mail posted in 1998, and a web page that he had  established.  This paper is offered as a resource so that De Witte's extraordinary discoveries  may be given the attention and study that they demand, and that others may be motivated to repeat the experiment, for that is the hallmark of science\footnote{The author has been developing and testing  new techniques for doing one-way RF travel time experiments.}.

\section{The De Witte Experiment}

In a 1991  research project  within Belgacom,
the Belgium telecommunications company,  another (serendipitous) detection of absolute motion was performed.  The study was undertaken by  Roland De Witte.   This organisation had two sets of atomic clock\index{atomic clocks}s in two buildings in Brussels separated by 1.5 km and the research project  was an investigation of  the task of synchronising these two clusters of atomic clocks. To that end  5MHz radio frequency (RF) signals were sent  in both
directions   through two  buried  coaxial cables linking the two clusters.   The atomic clocks were cesium beam atomic clocks, and there were three in each cluster: A1, A2 and A3 in one cluster, and B1, B2, and B3 at the other cluster. In that way the stability of the clocks could be established and monitored. One cluster was in a building on Rue du Marais and the second cluster
was due south in a building on Rue de la Paille.  Digital phase comparators were used to measure
changes in times between clocks within the same cluster and also in the propagation times of the RF
signals. Time differences between clocks within the same cluster showed  a linear phase drift caused
by the clocks not having exactly the same frequency, together with short term and long term noise.
However the long term drift was very linear and reproducible, and that drift could be allowed for
in analysing time differences in the propagation times between the clusters. 

The atomic clocks (OSA 312) and the digital phase comparators  (OS5560 ) were manufactured by Oscilloquartz, Neuch‰tel, Switzerland. The phase comparators produce a change of 1 V for a phase variation of 200 ns  between the two input signals.  At both locations the comparison between local clocks, A1-A2 and A1-A3, and between B1-B2, B1-B3, yielded linear phase variations in agreement with the fact that the clocks have not exactly the same frequencies due to the limited reproducible accuracy together with a short term and long term phase noise (A.O. McCoubrey, Proc. of the IEEE, Vol 55, No 6, June, 1967, pp. 805-814 ). 
Even if the long term frequency instability  were  $2 \times10^{-13}$ this is  able to produce a phase shift of 17 ns a day, but this instability was not often  observed and the ouputs of the phase comparators have shown that the local instability was typically only a few nanoseconds a day (5 ns) between two local clocks.

But  between distant clocks A1 toward B1 and B1 toward A1, in addition to the same linear phase variations (but with identical positive and negative slopes, because if one is fast, the other is slow), there is also an additional clear sinusoidal-like phase undulation ($\approx$ 24 h period) of the order of 28 ns peak to peak.

The possible instability of the coaxial lines cannot be responsible for  the phase effects observed because these signals are in phase opposition and also because the lines are  identical (same place, length, temperature, etc...) causing the cancellation of any such   instabilities.
As well the experiment was  performed over 178 days, making it possible to measure with accuracy ($\pm$ 25 s) the period of the phase signal to be the sidereal day (23 h 56 min ), thus permitting to conclude that  absolute motion had been detected in contradiction with the Einsteinian ``principle of relativity'', even with  apparent turbulence.

According to the manufacturer of the clocks, the typical humidity sensitivity is
$df/f = 10^{-14}$/\%humidity, so the effect observed between two distant clocks (24 ns in 12 h)  needs, for example, a differential step of variation of humidity of 55\%, two times a day, over 178 days. So the  humidity variations cannot be responsible for the persistent periodic phase shift observed.
As for pressure effects, the manufacturer confirmed that no measurable frequency change during pressure variations around 760 mm Hg had been observed.
When temperature effects are   considered, the typical sensitivity around room temperature is $df/f =0.25\times10^{-13}/^0$C and implies, for example, a differential step of room temperature variation of 24$^0$C, two times a day, over 178 days to produce the observed time variations. Moreover the room temperature was maintained at nearly a constant around 20$^0$C by the thermostats of the buildings. So the possible temperature variations of the clocks could not be responsible for the periodic phase shift observed between distant clocks. As well the heat capacity of the housings of the clocks  would even further smooth out  possible temperature variations.
Finally, the typical magnetic sensitivity of $df/f=1.4\times 10^{-13}$/Gauss needs, for example, differential steps of field induction of  4 Gauss variation, two times a day, over 178 days. But the terrestrial magnetic induction in Belgium is only in the order of 0.2 Gauss and thus its variations are much less (except during a possible magnetic storm). As for  possible parasitic variable DC currents in the vicinity of the clocks, a 4 Gauss change  needs a variation of 2000 amperes in a conductor at 1 m,  and thus can be excluded as a possible effect.
So temperature, pressure, humidity and magnetic induction effects on the frequencies of the clocks were thus  completely negligible in the  experiment.

\begin{figure}[t]
\hspace{20mm}\includegraphics[scale=0.35]{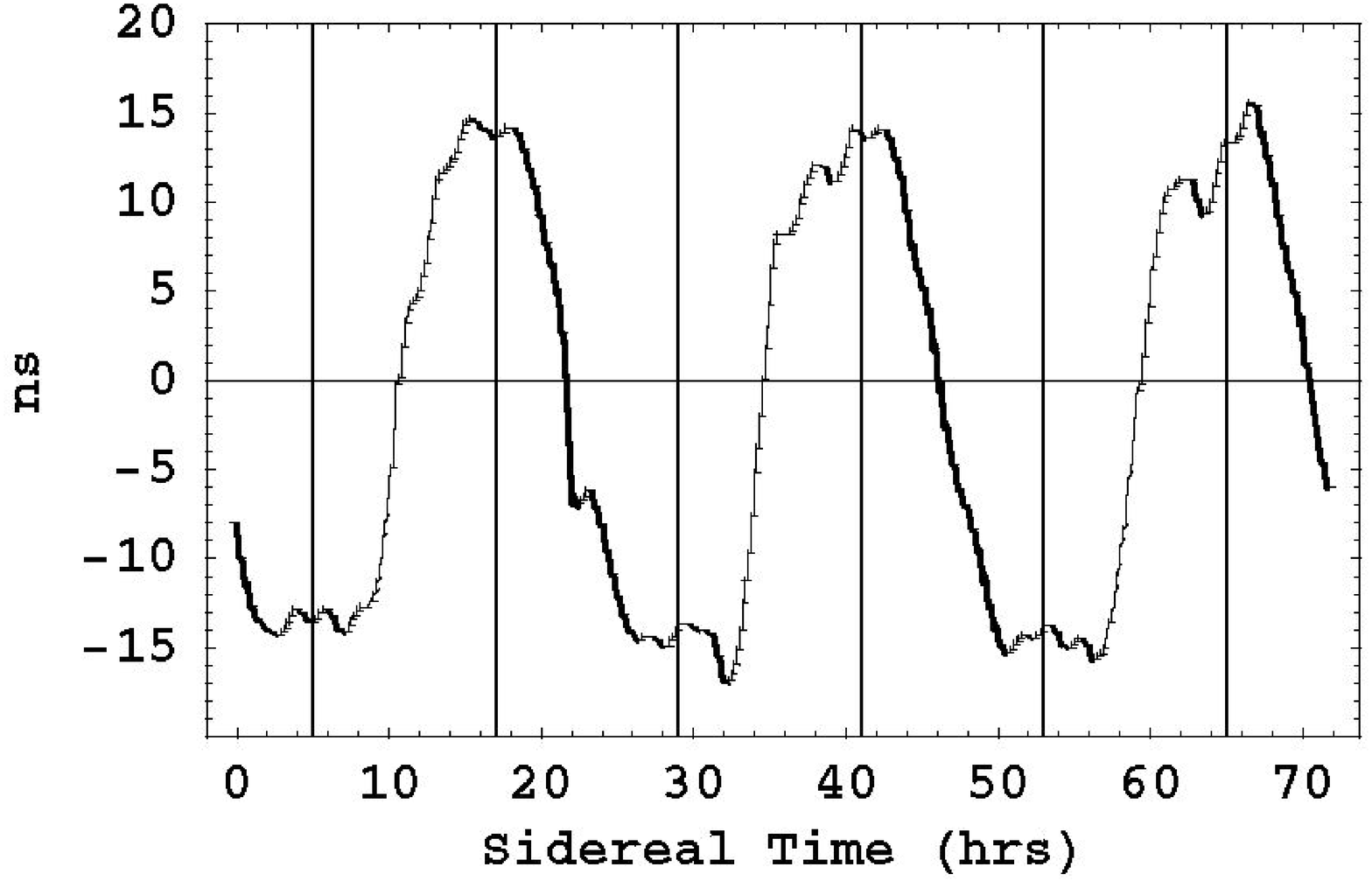}
\caption{\small{ Variations in twice the one-way travel time, in ns, for an RF signal to travel 1.5
km through a buried coaxial cable between  Rue du Marais and Rue de 
la Paille, Brussels, by subtracting the Paille Street phase shift data from the Marais Street  phase shift data. An offset  has been used  such that the average is zero.   The cable has a
North-South  orientation, and the data is $\pm$ difference of the travel times  for NS and SN
propagation.  The sidereal time for maximum  effect of $\sim\!\!5$hr (or   $\sim\!\!17$hr) (indicated
by vertical lines) agrees with the direction found by Miller \cite{Miller}. Plot shows
data over 3 sidereal days  and is plotted against sidereal time. The main effect is caused by the rotation of the earth. The superimposed fluctuations are evidence of turbulence i.e gravitational
waves. Removing the earth induced rotation effect we obtain the first experimental data of the turbulent structure of space, and is shown in Fig.\ref{fig:fractal}. De Witte performed this experiment over 178 days, and demonstrated that the effect tracked sidereal time and not solar time, as shown in Fig.\ref{fig:DeWitteST} }  
\label{fig:DeWittetimes}}\end{figure}

Changes in propagation times  were observed  over  178 days  from June 3 1991 7h 19m GMT  to 27 Nov 19h 47m GMT and recorded. A sample of the  data, plotted against sidereal time\index{sidereal time} for just  three days, is shown in Fig.\ref{fig:DeWittetimes}.  De Witte recognised that the data was evidence of absolute motion but he was unaware of the Miller experiment  and did not realise that the Right Ascension for minimum/maximum  propagation time agreed almost exactly with Miller's direction ($\alpha=5.2^{hr}, \delta=-67^0$). In fact De Witte expected that the direction of absolute motion should have been in the CMB direction, but that would have given the data a totally different sidereal time signature, namely the times for maximum/minimum would have been shifted by 6 hrs.  The declination of the velocity observed in this De Witte experiment cannot be determined from the data as only three days of data are available.   However assuming exactly the same declination as Miller  the speed observed by De Witte appears to be also  in excellent agreement with the Miller speed, which in turn is in agreement with that from the Michelson-Morley and other  experiments.

\begin{figure}
\hspace{20mm}\includegraphics[scale=0.35]{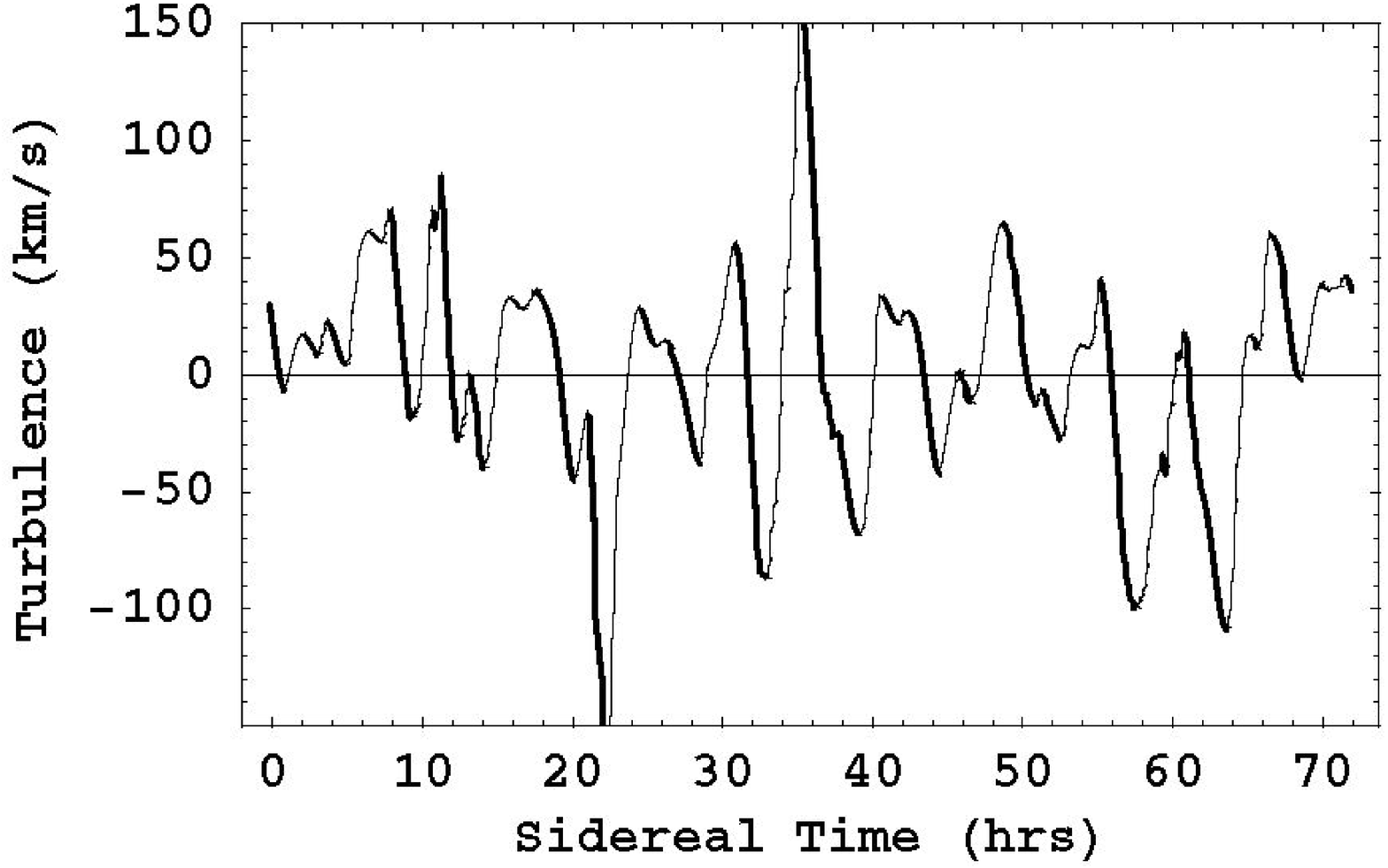}
\caption{\small{Shows the speed fluctuations, essentially `gravitational waves' observed by 	De Witte in 1991 from the measurement of variations in the RF coaxial-cable travel times.  This data is obtained from that in Fig.\ref{fig:DeWittetimes} after removal of the dominant effect caused by the rotation of the earth.  Ideally the velocity fluctuations are three-dimensional, but the De Witte experiment had only one  arm. This plot is suggestive of a fractal structure to the velocity field. This is confirmed by the
power law analysis  shown in  Fig.\ref{fig:powerlaw}. From \cite{Schrod}.}
\label{fig:fractal}}\end{figure}

\begin{figure}[t]
\hspace{30mm}\includegraphics[scale=1.5]{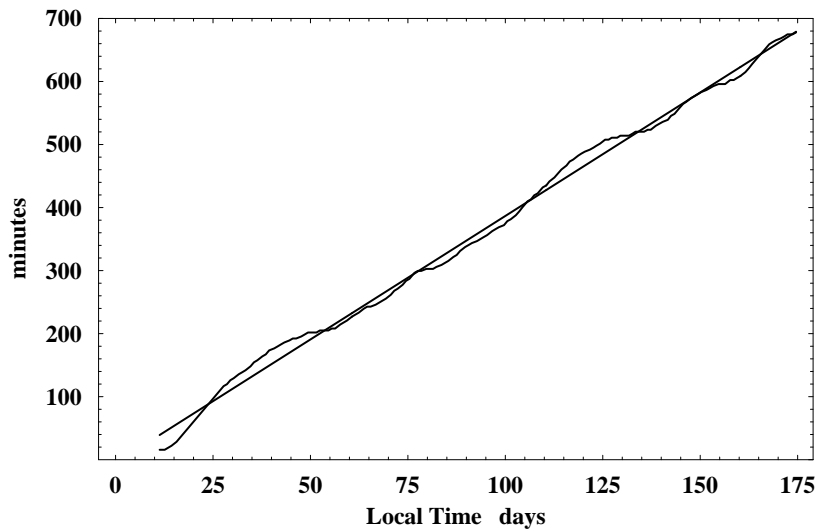}
\caption{\small{  Plot of the negative of the drift of the cross-over time between minimum and
maximum travel-time variation each day (at $\sim10^h\pm1^h$ ST) versus local solar time for some
178 days, from June 3 1991 7h 19m GMT  to 27 Nov 19h 47m GMT. The straight line plot is the least squares fit to the experimental data,  giving an average slope of 3.92 minutes/day. The time difference between a sidereal day and a solar
day is 3.93 minutes/day.    This demonstrates that the effect is related to sidereal time and not local
solar time. }
\label{fig:DeWitteST}}\end{figure}

\begin{figure}[t]
\hspace{30mm}\includegraphics[scale=0.30]{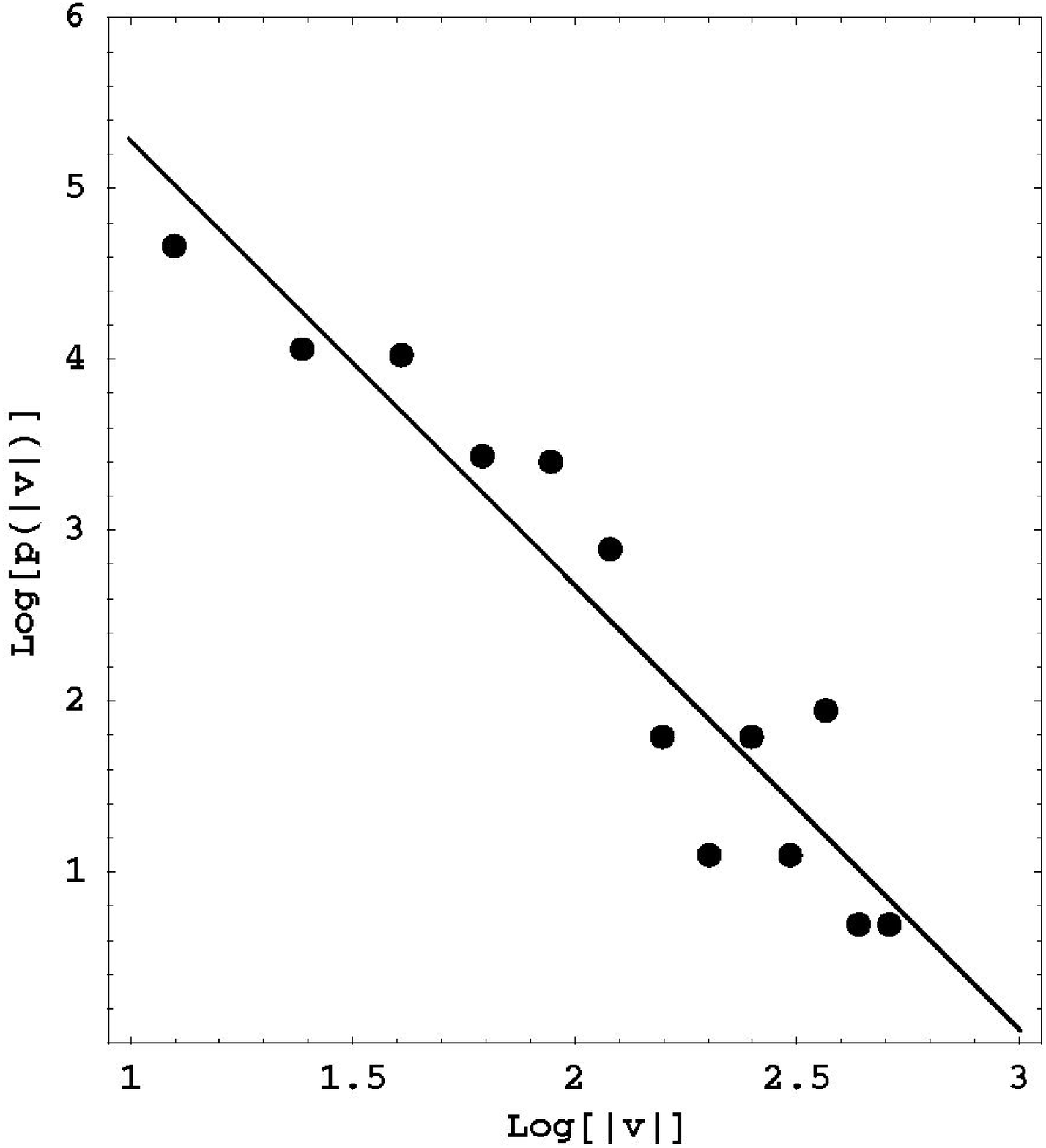}
\caption{\small{Shows that the speed  fluctuations in  Fig.\ref{fig:fractal} are scale free, as  the
probability distribution from binning the speeds has the form $p(v) \propto |v|^{-2.6}$. This plot shows
$Log[p(v)]$ vs $|v|$. From \cite{Schrod}.}
\label{fig:powerlaw}}\end{figure}

Being  1st-order in $v/c$  the Belgacom experiment is easily analysed to sufficient accuracy by
ignoring relativistic effects, which are 2nd-order in $v/c$.   Let the projection of the absolute
velocity vector ${\bf v}$ onto the direction of the coaxial cable be $v_P$.  Then the
phase comparators reveal the {\it difference}  between the propagation
times in NS and SN directions. Consider a simple analysis to establish the magnitude of the observed speed.
\begin{eqnarray}
\Delta t &=& \frac{L}{\displaystyle{\frac{c}{n}}-v_P}-
\frac{L}{\displaystyle{\frac{c}{n}}+v_P},\nonumber\\
&=& 2\frac{L}{c/n}n\frac{v_P}{c}+O(\frac{v_P^2}{c^2}) \approx 2t_0n\frac{v_P}{c}.
\label{eqn:DW1}\end{eqnarray}
Here $L=1.5$ km is the length of the coaxial cable, $n=1.5$ is the assumed refractive index of the insulator within
the coaxial cable, so that the speed of the RF signals is approximately $c/n=200,000$km/s, and so
$t_0=nL/c=7.5\times 10^{-6}$ sec is the one-way RF travel time when
$v_P=0$.  Then, for example, a  value of  $v_P=400$km/s would give $\Delta t = 30$ns. De Witte reported a speed of 500km/s.  Because Brussels has a latitude of $51^0$ N then for the Miller direction the projection effect is such that $v_P$ almost varies from zero to a maximum value of $|{\bf v}|$.  The De Witte  data in  Fig.\ref{fig:DeWittetimes} shows $\Delta t$ plotted with a false zero, but  shows a variation of some 28 ns.  So the De Witte data is in excellent  agreement with the Miller's data.

The actual days of the data in 
Fig.\ref{fig:DeWittetimes} are not revealed by De Witte so a detailed analysis of the  data
is not possible.   If all of De Witte's 178 days of data were available then a detailed analysis would be possible. 

De Witte does however reveal the sidereal time of the cross-over time, that is a `zero' time
in Fig.\ref{fig:DeWittetimes}, for all 178 days of data.  This is plotted in Fig.\ref{fig:DeWitteST} and
demonstrates that the time variations are correlated with sidereal time\index{sidereal time} and not local solar time\index{solar time}.  A least squares best fit of a linear relation to that data gives that the cross-over time is retarded, on average, by 3.92 minutes per solar day. This is to be compared with the fact that a sidereal day is 3.93 minutes shorter than a solar day. So the effect is certainly  galactic  and not associated with any daily thermal effects, which in any case would be very small as the cable is buried.  Miller had also compared his data against sidereal time and established the same property, namely that, up to  small diurnal effects  identifiable with the  earth's orbital  motion,  the dominant features in the data tracked sidereal time and not solar time, \cite{Miller}.

The De Witte data is also capable of resolving the question of the absolute direction of motion found by Miller. Is the direction ($\alpha=5.2^{hr}, \delta=-67^0$) or the opposite direction? Being a 2nd-order Michelson interferometer experiment Miller had to rely on the earth's orbital effects in order to  resolve this ambiguity, but his analysis of course did not take account of the gravitational in-flow effect \cite{Book, AMGE}.  The De Witte experiment could easily resolve this ambiguity by simply noting the sign of $\Delta t$.  Unfortunately it is unclear   as to how the sign in Fig.\ref{fig:DeWittetimes} is actually defined, and
De Witte does not report a direction expecting, as he did, that the direction should have been the same as the CMB direction.

The dominant effect in Fig.\ref{fig:DeWittetimes} is caused by the rotation of the earth, namely that the orientation of the coaxial cable with respect to the direction of the flow past the earth changes as the earth rotates. This effect may be approximately unfolded from the data, see \cite{Book, AMGE}, leaving  the gravitational waves shown in Fig.\ref{fig:fractal}.  This is the first  evidence that the velocity field describing  the flow of space has a complex structure, and is indeed fractal. The fractal structure, i.e. that there is an intrinsic lack of scale to these speed fluctuations, is demonstrated by binning the absolute speeds $|v|$ and counting the number of speeds $p(|v|)$ within each bin.  Plotting  Log$[p(|v|)]$ vs $|v|$, as shown  in Fig.\ref{fig:powerlaw} we see that $p(v) \propto |v|^{-2.6}$.  The Miller data also shows evidence of turbulence of the same magnitude.  So  far the data from three experiments, namely Miller, Torr and Kolen, and De Witte, show  turbulence in the flow of space past the earth.  This is what can be called gravitational waves \cite{Book, AMGE}.

\section{Biography of  De Witte}

Roland De Witte was born   September 29, 1953  in the small village of Halanzy in the  south of Belgium\footnote{These short notes were extracted from De Witte's webpage. }.
He became the apprentice to an electrician and learned electrical wiring of  houses. At the age of fourteen he decided to take private correspondence courses in electronics  from the EURELEC company, and obtained  a diploma at the age of sixteen. 
 He decided to stop  work as an apprentice and go to school. Without a state diploma it was impossible for him to be admitted into an ordinary school with teenagers of his age. After  working for a scrap company where he used dynamite, he was  finally admitted into a secondary school with the assistance  of the director, but with the condition that he pass some tests from the board of the state examiners, called the Central Jury, for the first three years. After having sat the exams he became a legitimate schoolboy. But when he was in the last but one year in secondary school he decided to prepare for the entrance exam in physics at  the University of Li\`{e}ge, and   became a university student in physics one year before his friends. During secondary school years he was interested in all the scientific activities and became a schoolboy president of the Scientific Youths of the school in Virton. Simple physics experiments were performed: Millikan, photoelectric effect, spectroscopy, etc... and a small electronics laboratory was  started.  He  also took part in different scientific short talks contests,  and became a prizewinner for a talk about  ``special relativity'', and  received a prize from the Belgian Shell Company  which had organised the contest. De Witte even  visited the house where Einstein lived for a few months in Belgium when he left Germany. 
The house is the ``Villa Savoyarde'' at ``Coq-Sur-Mer''  Belgium, and is just 200 m from the North Sea.
During secondary school  De Witte had hobbies such as astronomy and pirate radio transmission on 27 Mhz with a hand-made transmitter, with his best long distance communication  being with Denmark. 

De Witte says  that he is not able to study by ``heart'', and  during secondary school, even with his bad memory which caused problems in history and english,  he nevertheless  always achieved the maximum of points in physics, chemistry and mathematics and was the top of his class.
At University he obtained the diploma from the   two year degree in physics but was not able to continue due to the ``impossibility to study by heart several thousands of pages of erroneous calculations''  like the others did to obtain the graduate  diploma. Thus even though considered to be intelligent by several teachers, he decided to leave the University and  became the manager of a retail  electronic components shop. He did this job for ten years while also  performing  his physics experiments and studying theoretical physics. He was interested in microwaves and became an  IEEE member and reader of the publications of the Microwave Theory \& Techniques and Instrumentation \& Measurement Societies. During that period he built an electron spin resonance spectrometer  for the pleasure of studying the electron and free radicals. By chance he was invited by  Dr. Yves Lion of the Physics Institute of the University of Li\`{e}ge to help them for a few weeks in their researches on the photoionisation mechanism of the tryptophan amino-acid with the powerful EPR spectrometer. 
He was also interested in TV satellite reception and Meteosat images. He  built several microwave microstrip circuits such as  an 18 dB low noise amplifier using GaAs-Fets for 11.34 GHz. He also developed some apparatus using microprocessors for a digital storage system for Meteosat's images.

In 1990 he became a civil servant in the Metrology Department of the Transmission Laboratories of Belgacom (Belgium Telephone Company). His job was to test the synchronization of rubidium frequency standards on a distant master ceasium beam clock. It is there that he took the time to compare the phase of distant ceasium clocks and discovered the periodic phase shift signal  with a sidereal day period.  De Witte  retired from the Department, reporting that he had  been dismissed, and worked on theoretical physics and  philosophy of science, while performing various cheap experiments to test his  electron theory and also develop a new working process for a beamless  ceasium clock.

De Witte acknowledged  assistance from  J. Tamborijn, the Engineer Cerfontaine, and particularly   Engineer and Executive Director B. Daspremont, all from the  Metrology, Fiber Optics and Transmission Laboratory of Belgacom in Brussels, for the use of the six caesium atomic clocks, the comparators, the recorder and the underground lines, and also    Paul P\`{a}quet, Director of the Royal Observatory of Belgium, for explanations and   documentation provided about the realisation of UTC in Belgium.

\section{De Witte's Publication}

Roland De Witte was not able to have his experimental results published in a physics journal.  His only known publications are that of an e-mail posted to the newsgroup sci.physics.research, and his webpage.  The e-mail is reproduced here:

\begin{quote}{\it Ether-wind detected!

    * Subject: Ether-wind detected!
    
    * From: "DE WITTE Roland" \newline \hspace{25mm} $<$roland.dewitte@ping.be$>$
    
    * Date: 07 Dec 1998 00:00:00 GMT
    
    * Approved: baez@math.ucr.edu
    
    * Newsgroups: sci.physics.research
    
    * Organization: EUnet Belgium, Leuven, Belgium

I have performed an interesting experiment with cesium beam frequency
standards.

A 5 Mhz signal from one clock (A ) is sent to another clock (B) 1.5 km apart
in Brussels by the use of an underground coaxial cable of the Belgium
Telephone Company.  There, the 5Mhz signal from clock A is compared to 
the one of clock B, by the use of a digital phase comparator (like those 
used in PLL).

Incredibly, the output of the phase comparator shows a clear and important
sinus-like undulation which permits to conclude of the existence of a
periodic variation (24 h period)) of the speed of light in the coaxial cable
around 500 km/s.

In performing the experiment during 178 days, with six cesium beam clocks,
the period of the phase signal has been accurately measured and is 23h 56 m
+- 25 s. and thus is the sidereal day.

This result, like the one of D.G. Torr and P. Kolen (Natl. Bur. Stand.
(U.S.), Spec. Publ. 617, 1984) is well understood with a new space-time
theory based on a new electron theory.

It is also the case for the nearly negative result of the experiment of
Krisher et al, with a fiber optics instead of a coaxial cable (Physical
review D, Vol 42, number 2, 1990, pp. 731-734).

All the details of the experiment is on my web-site under construction:
www.ping.be \newline / electron/belgacom.htm together with already a few arguments
against Einstein's special theory of relativity.

DE WITTE Roland

www.ping.be/electron

[Moderator's note: needless to say, there are many potential causes of
daily variations that need to be studied in interpreting an experiment 
of this sort. - jb]}
\end{quote}

\section{Conclusions\label{section:conclusions}}

The De Witte experiment was  truly remarkable considering that initially it was  serendipitous.  The data demonstrated yet again that the Einstein postulates were in contradiction with experiment.   No physics journal has published a report from  De Witte, although he did make a submission  for publication to the Annals of the Louis de Broglie's Foundation.  De Witte himself reported that he was dismissed from Belgacom. Papers reporting or analysing  absolute motion and related effects continue to be  banned by  mainstream physics journals. This appears to be based on the almost universal misunderstanding by physicists that absolute motion is incompatible with the many confirmed relativistic effects.  DeWiite's  data like that of Miller is extremely valuable and needs to be made available for detailed analysis. Regrettably Roland De Witte has died, and  the bulk of the data was apparently lost when he left Belgacom.

This work is supported by an Australian Research Council Discovery Grant.

\end{document}